\documentclass[letterpaper, 10 pt, conference]{ieeeconf}

\usepackage[utf8]{inputenc}
\usepackage[english]{babel}
\usepackage[normalem]{ulem}
\usepackage[T1]{fontenc}
\usepackage{url,comment}

\usepackage{graphicx}

\usepackage[margin=1in]{geometry}
\usepackage{amssymb,amsmath,bm}
\usepackage{mathtools}
\usepackage{mathptmx}

\usepackage{footmisc}

\newcommand{\V}{{\cal V}}
\newcommand{\E}{{\cal E}}

\newcommand{\x}{{\bm x}}

\newcommand{\w}{{\bm w}}

\usepackage[dvipsnames]{xcolor}
\usepackage{xspace}

\title{Universality and Control of Fat Tails
}

\author{
Michael (Misha) Chertkov [chertkov@arizona.edu] 
\\
GIDP in Applied Mathematics \& Department of Mathematics, University of Arizona, Tucson, AZ}

\date{\today}

\begin{document}
\bstctlcite{IEEEexample:BSTcontrol}

\maketitle

\begin{abstract}
Motivated by applications in hydrodynamics and networks of thermostatically-control loads in buildings we study control of linear dynamical systems driven by additive and also multiplicative noise of a general position. Utilizing mathematical theory of stochastic multiplicative processes we present a universal way to  estimate fat, algebraic tails of the state vector probability distributions.  This  prompts us to introduce and analyze  mean-$q$-power stability criterion, generalizing the mean-square stability criterion, and then juxtapose it to other tools in control.
\end{abstract}

\section{Introduction}\label{sec:intro}

Study of multiplicative noise models has a long  history in control,  with many foundational results reported in late  20th century and early 21st century  \cite{kats_stability_1960,kushner_stochastic_1967,wonham_optimal_1967,kozin_survey_1969,kleinman_optimal_1969,mclane_optimal_1971,willems_feedback_1976,bernstein_robust_1987,
boyd_linear_1994,
de_oliveira_state_2001}, 
\footnote{See also a review-like recent paper \cite{gravell_robust_2020} and references therein.}. 
These classic approaches included testing stability of the control solutions via perturbations, analyzing the mean-square stability measure \cite{kleinman_optimal_1969,boyd_linear_1994}, 
and also studying the multiplicative version of the Linear Quadratic Gaussian (LQG) control problem by solving the generalized Ricatti equations \cite{wonham_optimal_1967,mclane_optimal_1971}, then followed by  a path to efficient resolution via semi-definite-programming computations of the linear inequality type \cite{boyd_linear_1994,el_ghaoui_state-feedback_1995}. 

\begin{figure}[h]
    \centering
    \vspace{-0.4cm}
    \includegraphics[width=0.9\linewidth]{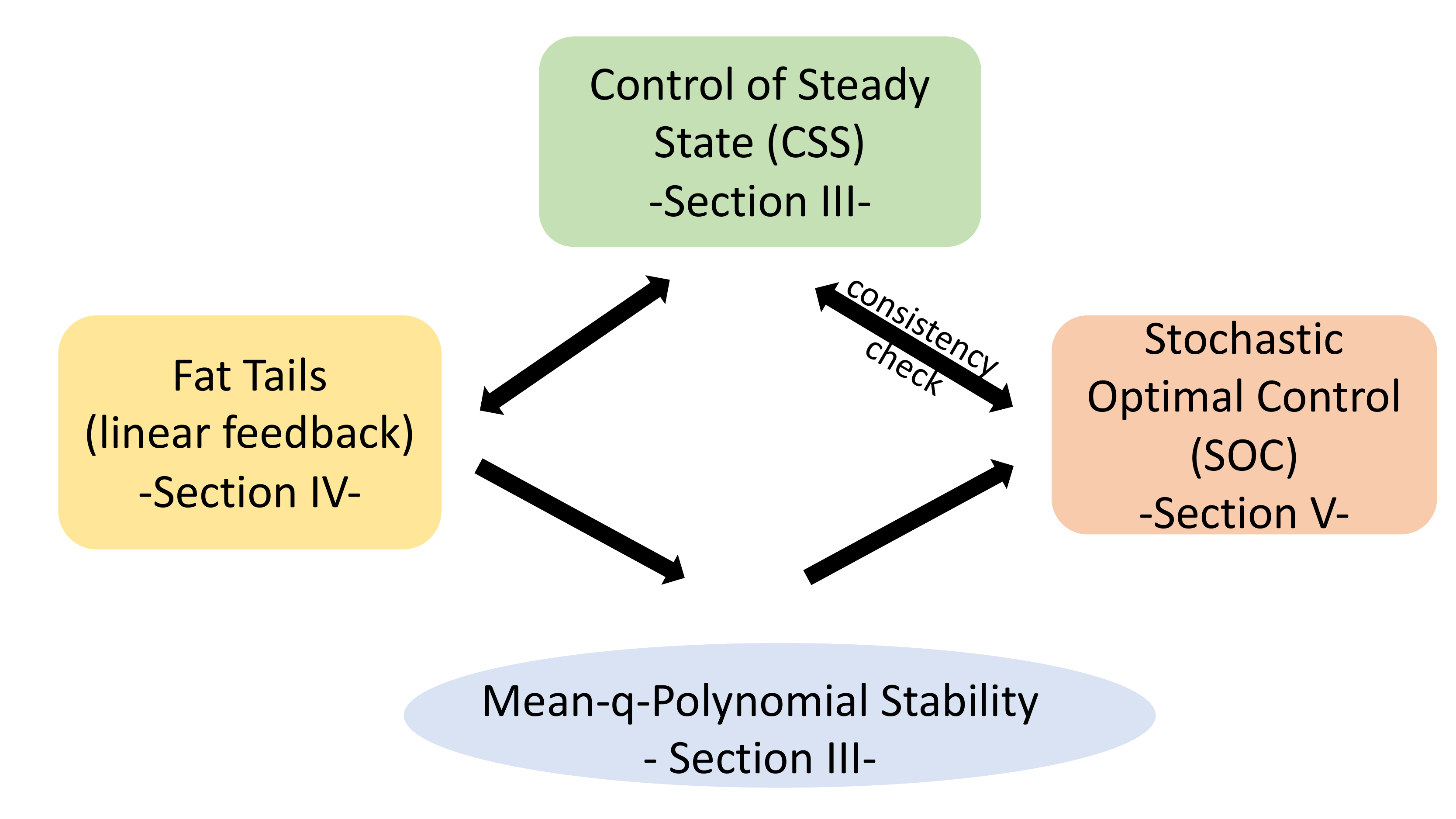}%
    \caption{Section-to-Concepts map of the letter. \vspace{-1cm}}
    \label{fig:road-map}\vspace{-0.2cm}
\end{figure}

This letter, inspired by applications in hydrodynamics and thermal control of buildings, and also by theoretical overlap with studies in stochastic multiplicative processes \cite{ruelle_ergodic_1979,goldhirsch_stability_1987} and in statistical hydrodynamics \cite{ 
falkovich_particles_2001} suggests the following complementary (and to the best of our knowledge novel) contribution to the classic subject of stability and control of the multiplicative linear systems:\\ 
(1) We introduce and analyze effect of a multiplicative noise of a general position,  that is possibly not white and non-Gaussian, on stability of a linear state feedback solution. Utilizing the theory of  stochastic multiplicative processes, and specifically the so-called Oseledets theorem
\cite{ruelle_ergodic_1979,goldhirsch_stability_1987,falkovich_particles_2001}, we observe that if the probability distribution of the state vector stabilizes it shows fat, algebraic tail  with the exponent which scales linearly with the vector of the feedback rates. Moreover, we express the algebraic exponent via  universal characteristics of the so-called time-ordered exponential of the multiplicative matrix, in particular via Cram\'{e}r function of the Lyapunov exponents measuring the exponential rates of growth of uncertainty in different components of the initial state vector. Our analysis also reveals that at the times larger than inverse of the largest Lyapunov exponent ANY multiplicative noise can be described by a white-Gaussian substitute with a properly re-normalized covariance.\\
(2) Motivated by (1) we introduce the Mean-$q$-Power (M$q$P) stability criterion, requiring that the mean of the $q/2$-th moment of the state-vector squared $\lim_{t\to\infty}\mathbb{E}[(\x\x^T)^{q/2}]$ is finite. (This criterion is a generalization of the standard in control theory mean-square stability criterion, correspondent to the case of $q=2$.) 
The M$q$P criterion is useful because flexibility in the choice of the parameter $q$ helps us to better test stability of the feedback control in the cases where it results in the algebraic tails of the state vector probability distribution function.
Exploring the regime of the ``slower than inverse of the largest Lyapunov exponent" control, and thus replacing multiplicative noise of a general position by white-Gaussian, we are able, following classic approaches of the control theory,  to estimate minimal linear-state feedback which guarantees the M$q$P-stability and then verify consistency with the long time asymptotic of the noise-multiplicative version of the Linear Quadratic Gaussian approach.

Sections-to-Concept map of the letter is shown in Fig.~(\ref{fig:road-map}). Reader interested in theory (only) is advised to check the main part of Sections \ref{sec:basic}, \ref{sec:CSS}, skipping Subsection devoted to the applications, and then see Section \ref{sec:synthesis} where synthesis of the main general results, concerning the probability distribution of the state vector in the controled/stabilized regime, are reported. The two applications -- ``swimmers" and ``thermal" -- are introduced in Section \ref{sec:swimmers}  and Section \ref{sec:building} and the results are presented in Sections \ref{sec:swimmers-CSS},\ref{sec:swimmers-SOC} and Sections \ref{sec:thermal-CSS},\ref{sec:thermal-CSS-multi},\ref{sec:thermal-SOC}, respectively.

\section{Basic Dynamic Model}\label{sec:basic}

Consider stochastic dynamics of a state vector ${\bm x}(t)\in \mathbb{R}^d$  which is governed by the following linear equation:
\begin{gather}\label{eq:SDE}
    \frac{dx_i}{dt}=\sum\limits_j \left(m_{ij}+\sigma_{ij}(t)\right) x_j(t) +\xi_i(t) +u_i(t),
\end{gather}
where $i=1,\cdots,d$ and ${\bm m}= (m_{ij}:i,j=1,\cdots,d)$ is a constant matrix; ${\bm \xi}(t)= (\xi_i(t):i=1,\cdots, d)$ is a zero-mean white-Gaussian noise  fully described by 
\begin{gather*} 
\forall i,j:\quad \mathbb{E}\left[\xi_i(t)\xi_j(t')\right]=\kappa_i \delta_{ij}\delta(t-t');
\end{gather*}
${\bm u}(t)=(u_i(t):i=1,\cdots,d)$ is a vector of control; and the multiplicative matrix ${\bm \sigma}(t)=(\sigma_{ij}(t):i,j=1,\cdots,d)$ is a zero-mean stochastic and independent of the vector of additive noise ${\bm \xi}$. We consider two ways to model ${\bm \sigma}$ in Eq.~(\ref{eq:SDE}) -- {\it special} and {\it general}. In the {\it special} case ${\bm\sigma}$  is white-Gaussian  with a constant covariance. To introduce the {\it general} model of ${\bm\sigma}(t)$ we consider an auxiliary multiplicative dynamics
\begin{gather}\label{eq:W}
     \frac{d}{dt}{\bm W}={\bm \sigma} {\bm W},
\end{gather}
where the matrix ${\bm W}\in \mathbb{R}^d\times \mathbb{R}^d$ is called the time-ordered exponential of ${\bm \sigma}$.  According to the Oseledets theorem
(see \cite{ruelle_ergodic_1979,goldhirsch_stability_1987,falkovich_particles_2001} and references there in) at sufficiently large  times $t$ the matrix
$\log ({\bm W}^+{\bm W})/t$ stabilizes. That is eigenvectors of the matrix tend to $d$ fixed orthonormal eigenvectors ${\bm f}_i$ of ${\bm W}$ and  the respective set of ordered eigenvalues $\lambda_i = \log |{\bm W} {\bm f}_i|/t$, where $\lambda_1\geq \lambda_2\geq \cdots \lambda_d$ called the Lyapunov exponents, stabilize to their mean-values asymptotically:
\begin{gather}\label{eq:Cramer}
    \hspace{-0.5cm}\text{\underline{general}}:\ P(\lambda_1,\cdots, \lambda_d|t)\propto \exp\left(-t S(\lambda_1,\cdots,\lambda_d)\right), 
\end{gather}
where $S(\cdot)$ is the so-called Cr\'amer function. 

In the remainder of this Section we present two examples where the basic model applies.

\subsection{Active and Passive Swimmers}\label{sec:swimmers}

We consider a smooth, chaotic velocity field of a general position discussed extensively in stochastic hydrodynamics,  see 
review \cite{falkovich_particles_2001} and references therein. Particles/swimmers which are placed in such a flow separate exponentially fast. Our task is to navigate the active swimmer to control its separation from the passive swimmer, assuming that the two were released at the same position initially. The vector of separation of the two swimmers ${\bm r}=(r_i:i=1,\cdots,d)$ in $d$ dimensions (where $d=2,3$) evolves according to 
\begin{gather}\label{eq:swim-r}
    -\alpha \Big(\frac{d{\bm r}}{dt}-{\bm \sigma}(t) {\bm r}\Big)= {\bm u}(t)+ {\bm  {\xi}}(t),
\end{gather}
where $\alpha$ is the friction coefficient (which we set to unity without loss of generality) and ${\bm u}$ and ${\bm  {\xi}}$ are control force exerted by the active swimmer and the difference of the thermal forces acting on the active and thermal swimmers, respectively. The ${\bm \sigma}(t) {\bm r}$ term in Eq.~(\ref{eq:swim-r}) represents the first term of the Taylor expansion of the velocity difference between the two swimmers in the separation vector between the two. (This expansion is justified in the case of a smooth large scale flow.) Following assumptions which are standard in stochastic hydrodynamics \cite{falkovich_particles_2001} we will model ${\bm \sigma}(t)$ and ${\bm  {\xi}}$ as stochastic and independent.

\subsection{Dynamics of Temperature in Multi-Zone Buildings}\label{sec:building}

We follow a ``gray box" modeling of the multi-zone building describing thermal exchange between a zone and outside environment and between a zone and the building's Air Handling Unit (AHU), as  discussed in \cite{valenzuela_statistical_2023} (see also references therein), and then generalize the model to account for thermal exchange between neighboring zones \cite{picard_impact_2017}. Consider, first, the case of a single zone, where temperature within the zone is governed by
\begin{gather}\label{eq:T}
\frac{d T}{dt}=-c_o(T-T_o)-c_s(T-T_s)u(t)+\xi(t),
\end{gather}
where $c_o$ and $c_s$ are rates of thermal exchange between the zone $i$ and the "outside" environment (which can be viewed as outside of the building, but may also represent an aggregation of other zones of the building) and the AHU, respectively (the rates are measured in the units of inverse time); $T_o$ and $T_s$ are ``outside" and AHU temperatures, respectively; $ \xi(t)$ is the white-Gaussian noise modeling additive fluctuations, with the covariance $\kappa$ proportional to occupancy and thus expressing behavioral uncertainty; and $u(t)$ is control of the opening in the pipe connecting zone $i$ to the AHU. Given fluctuations of the zone's occupancy in time, it is also reasonable to model its contribution to the thermal exchange with the "outside"  as split into the mean (constant or slowly dependent on time) and uncertain (thus stochastic) term: $c_{o}= \underline{c}_{o}+ \sigma(t)$. Now, suppose that $T_o$ and  $T_s$ are constant and assume that a single zone control $u(t)$ is split into constant and linear feedback components, i.e., $u(t)=\underline{u}+\phi\theta$, where $\theta=T-\underline{T}$ is deviation from the desired comfort temperature $\underline{T}$ and the constant component of the control $\bar{u}$ is chosen according to
$0=-\underline{c}_o(\underline{T}-T_o)-c_s(\underline{T}-T_s)\underline{u}$ thus guaranteeing that when $\xi$ and $\sigma$ are set to zero $T$ stabilizes to $\underline{T}$. Then Eqs.~(\ref{eq:T}) results in the following stochastic ODE for $\theta$
\begin{gather}\label{eq:theta}
\frac{d \theta}{dt}=-c(\phi)\theta+\tilde{\xi}(t)-\sigma(t)\theta,
\end{gather} 
where $c(\phi)= c_0+c_1\phi$, $c_0= \underline{c}_o+c_s\underline{u}$, $c_1= c_s(\underline{T}-T_s)$ and $\tilde{\xi}(t)= \xi(t)+T_o \sigma (t)$. 

Network generalization of Eq.~(\ref{eq:theta}) accounting for thermal flows between zones results in the following system of equations for the components of the temperature vector ${\bm \theta}=(\theta_i:i\in\V)$ (counted from the comfort temperature $\underline{T}$ set the same for all zones in the building), where $\V$ stands for the set of zones:
\begin{align} \label{eq:theta_i} 
\hspace{-0.3cm}\frac{d \theta_i}{dt}\!\!=\!\!-\!\left(\! c_i({\bm \phi})\!+\!\sigma_{io}\right)\theta_i 
\!-\!\!\!\!\!\!\!\!\!\!\sum\limits_{j:\{i,j\}\in \E}\!\!\!\!\!\!\!\!\!\!\left(\underline{c}_{ij}\!+\!\sigma_{ij}\!\right)(\theta_i\!-\!\theta_j) \!+\!\xi_i(t).
\end{align}
Here $\E$ denotes the set of edges in the network linking neighboring zones;  $c_{ij}$ is the rate of thermal exchange between the pair of neighboring zones $(i,j)\in\E$; it is assumed that the constant components of the control vector $\underline{\bm u}=(\underline{u}_i:i\in\V)$ are chosen according 
to $\forall i:\ 0=\underline{c}_{io}(\underline{T}-T_o)+c_{is}(\underline{T}-T_s)\underline{u}_i$, where $c_{io}$ and $c_{is}$ are the rates of thermal exchange, respectively, between zone $i$ and the outside environment, kept at the constant temperature $T_o$, and between zone $i$ and the AHU, kept at the constant temperature $T_s$; and $ c_i({\bm \phi})= \underline{c}_{io}+\sum_j c_{js}(\underline{T}-T_s)\phi_{ij}$, where ${\bm \phi}=(\phi_{ij}:i,j\in\V)$ is vector of the linear feedback rates. We also assume that both $c_{io}$ and $c_{ij}$ are split into constant and fluctuating parts,
$\forall i:\ c_i= \underline{c}_{io}+\sigma_{io};\ \forall (i,j):\ c_{ij}= \underline{c}_{ij}+ \sigma_{ij}(t)$.

It is clear that Eq.~(\ref{eq:swim-r}), with ${\bm r}$ substituted by ${\bm x}$, and Eqs.~(\ref{eq:theta},\ref{eq:theta_i}), with ${\bm \theta}$ substituted by ${\bm x}$ with properly re-defined uncertainty/noise terms, constitute particular cases of Eqs.~(\ref{eq:SDE}).

\section{Control of Steady State (CSS)} \label{sec:CSS}

Consider state feedback control, that is ${\bm u}(t)\to \w(\x(t))$,
where $\w(\cdot)$ is a yet-to-be-defined parameterized function. Then in the case of the white-Gaussian ${\bm \sigma}$ Eq.~(\ref{eq:SDE}) results in the following Kolmogorov-Fokker-Planck (KFP) equation for the stationary probability density function of the state vector ${\bm x}$ conditioned to $\w(\cdot)$:
\begin{align}\label{eq:KFP}
\hat{\cal L}P(\x|\w)=0,\quad \hat{\cal L} = &
\partial_{x_i}\left(w_i({\bm x})+m_{ij} x_j\right)\\ \nonumber  & + \kappa_{ij}\partial_{x_i}\partial_{x_j}+D_{ik;j\ell}\partial_{x_i}  \partial_{x_j} {\color{black} x_k} x_{\ell},
\end{align}
where $\hat{\cal L}$ is a second order differential operator and $\kappa$ and $D$ are elements of the matrix and tensor of covariances associated with the additive and multiplicative terms, respectively. 

Assuming that the steady state is achieved, i.e. that solution of the stationary KFP equation is well-defined, we pose the following
steady version of the stochastic optimal control 
\begin{align} \nonumber 
     &{\bm\phi}^*=\text{arg}\min\limits_{{\bm \phi}} \bar{C}({\bm\phi}),\quad \bar{C}({\bm\phi})=\int d\x P(\x|\w_{\bm \phi}) C(\x,\w_{\bm\phi}),\\ 
& C(\x,\w_{\phi}) =\underbrace{C_c(\w_{\phi})}_{\text{cost of control}}+\underbrace{C_g(\x)}_{\text{cost of achieving the goal}}, \label{eq:steady-control}
\end{align}
where $\phi$ stands for a vector of parameters selected to represent $\w_{\bm \phi}(\cdots)$.

Notice, that the CSS analysis will also help us to solve the M$q$P-stability problem.  Indeed, we will see below that the steady state (if settled) results in the algebraic decay of $P(\x|\w_{\bm \phi})$ with $|\x|$ and thus the $\phi$-parameterized linear state feedback is M$q$P-stable, i.e. the integral 
\begin{gather}\label{eq:C-q}
\int d\x P(\x|\w_{\bm \phi}) (\x \x^T)^{q/2}
\end{gather} 
is convergent, if $\phi$ is sufficiently large or if $q$ is sufficiently small. This explains our choice of the cost-of-achieving-the-goal in the M$q$P form $C_g(\x)\to \beta (\x \x^T)^{q/2}$ in the following Subsections,  where we discuss solution of the CSS problem on our two enabling examples.

\subsection{Swimmers CSS: short-correlated large-scale flow}\label{sec:swimmers-CSS}

Consider the swimmers' linear state feedback version of Eqs.~(\ref{eq:steady-control}),  thus with ${\bm x}$ substituted by ${\bm r}$,
\begin{gather}
    \label{eq:simple-swimmers} 
    \hspace{-0.3cm} {\bm u}(t)\to {\bm w}({\bm r})=\phi r,\  C_c\{\w\}\to \w^2,\ C_g({\bm r})\to \beta r^q,
\end{gather}
and apply to the stochastic dynamics governed by Eq.~(\ref{eq:swim-r}). Let us also choose Batchelor-Kraichnan model for the chaotic flow in $d$-dimensions  described by the following pair-correlation function of the velocity gradient matrix ${\bm \sigma}$ entering Eq.~(\ref{eq:swim-r}):
\begin{align}\label{eq:sigma-cov}
    & \forall i,j,k,l=1,\cdots,d:\  \mathbb{E}\left[\sigma_{ij}(t)\sigma_{kl}(t')\right]= \\ \nonumber & \hspace{0.5cm} D (d+1)\delta(t-t')\left(\delta_{jl}\delta_{ik}-\frac{\delta_{ij}\delta_{kl}+\delta_{jk}\delta_{il}}{d+1}\right),
\end{align}
where $\delta(\cdot)$ and $\delta_{ij}$ are the $\delta$-function and the Kronecker symbol respectively. Then,  the KFP Eq.~(\ref{eq:KFP}) for the spherically symmetric probability density 
$P(r|\phi)$, where $r=|{\bm r}|$, becomes
(see \cite{chertkov_how_1998} for details)
\begin{align}\label{eq:KFP-swimmers}
& {\cal L}_{sw}P(r|\phi)=0,\\ \nonumber & {\cal L}_{sw}=r^{1-d}\frac{d}{dr} r^d\left(\phi+\frac{1}{2}\left(D (d-1)r+\frac{\kappa}{r}\right)\frac{d}{dr}\right).
\end{align}
Here in Eq.~(\ref{eq:KFP-swimmers}) $\kappa$ stands for covariance of the thermal noise in Eq.~(\ref{eq:swim-r}). Solution of Eq.~(\ref{eq:KFP-swimmers}) is 
\begin{align}\label{eq:P-sw}
& P(r|\phi) =N^{-1}\left(\frac{\kappa}{D}+(d-1)r^2\right)^{-\phi/((d-1)D)},\\ \nonumber
& N = 2^{\phi/(d-1)} d \left(\frac{(d-1)D}{\kappa \pi}\right)^{d/2} \frac{\Gamma\left(\phi/(D(d-1))\right)}{\Gamma\left(\phi/(D(d-1))-d/2\right)},
\end{align}
where $N$ is the normalization coefficient which guarantees that $\int_0^\infty \Omega_r  dr P(r|\phi)=1$ and $\Omega_r=(\pi^{d/2}/\Gamma(d/2+1)) r^{d-1}$.
The solution is valid, i.e. the normalization integral is bounded, if $\phi> (d-1) d D/2$.

Substituting $P(r|\phi)$, given by Eq.~(\ref{eq:P-sw}), into Eq.~(\ref{eq:steady-control}) with $C_c(\cdots)$ and $C_g(\cdots)$ chosen according to Eq.~(\ref{eq:simple-swimmers}) we observe that the cost of control is well-defined, i.e. the respective integral converges and the control is M$q$P-stable at $\phi\geq \phi^{(s)}=(d+q)(d-1)/2$. To find optimal $\phi^{(*)}$, which should obviously be larger than the M$q$P value $\phi^{(*)}>\phi^{(s)}$, one needs to solve a straightforward one-parametric convex optimization. The solution, which is generally a bulky but explicit expression in terms of special function, simplifies in the $q=2$ case: 
\begin{align} 
    \phi^* = \frac{D(d\!+\!2)(d\!-\!1)\!+\!\sqrt{4\beta\!+\!D^2(d\!+\!2)^2(d\!-\!1)^2}}{2}.\label{eq:steady-swimmers-opt} 
\end{align}

\subsection{Thermal CSS: single zone}\label{sec:thermal-CSS}

Assuming that $\tilde{\xi}(t)$ and $\sigma(t)$ in Eq.~(\ref{eq:theta}) are zero-mean, independent and white-Gaussian  with the pair correlation functions described by
\begin{gather}\label{eq:-tilde-xi+sigma}
    \mathbb{E}\left[\tilde{\xi}(t)\tilde{\xi}(t')\right]/\kappa =\mathbb{E}\left[\sigma(t)\sigma(t')\right]/D=\delta(t-t'),
\end{gather}
we arrive at the following KFP equation for the probability distribution function of $\theta$ 
\begin{gather}\label{eq:KFP-thermal}
\left(\partial_\theta c(\phi)\theta +\kappa\partial^2_\theta +D{\color{black}\partial_\theta^2 \theta^2}\right)P(\theta|\phi)=0.
\end{gather} 
Solution of Eq.~(\ref{eq:KFP-thermal}), 
\begin{gather}\label{eq:thermal-stat}
    \vspace{-0.2cm} P(\theta|\phi)\!=\!\sqrt{\frac{D}{\pi\kappa}}{\color{black}\frac{\Gamma\left(\frac{c(\phi)}{2D}\right)}{\Gamma\left(\frac{c(\phi)}{2D}-\frac{1}{2}\right)}\left(\!1\!+\!\frac{D\theta^2}{\kappa}\!\right)^{-\frac{c(\phi)}{2D}}}\!,
\end{gather}
is normalizable if ${\color{black} c(\phi)>D}$. Then expectation of the cost evaluated according to Eqs.~(\ref{eq:steady-control})  is finite and convex at $c(\phi)=c_0+c_1\phi >{\color{black} D\max(q,2)}$, i.e. the system is M$q$P-stable at $\phi>\phi^{(s)}= {\color{black}(D\max(q,2)-c_0)}/c_1$. The optimal value is achieved at $\phi^* >\phi^{(s)}$ which returns the following explicit expression at $q=2$:
\begin{align}
    \phi^* =\frac{{\color{black}3}D-c_0+\sqrt{({\color{black}3}D-c_0)^2+\beta c_1^2}}{c_1}.
    \label{eq:phi*-thermo}
\end{align} 

\subsection{Thermal CSS: multi-zone}\label{sec:thermal-CSS-multi}

Assuming that $\xi_{i}$, $\sigma_{io}$ and $\sigma_{ij}$ entering Eq.~(\ref{eq:theta_i}) are independent, zero mean, white-Gaussian  with covariances $\kappa_{i}$, $D_{io}$ and $D_{ij}$, respectively, we arrive at the following multi-zone version of the single-zone KFP Eq.~(\ref{eq:KFP-thermal}): {\color{black}
\begin{align}\label{eq:KFP-multi}
&\hat{\cal L}_{m} P({\bm \theta}|{\bm \phi})\!=\!0,\ \hat{\cal L}_{m} \! = \!\!\!\!
\sum\limits_{i\in\V}\!\!\!  
\left(c_i\partial_{\theta_i} \theta_i+D_{io} \partial_{\theta_i}^2 \theta_i^2\!+\!\!\kappa_i\partial_{\theta_i}^2\right)\\ \nonumber & +\!\!\!\!\!\!\!\sum\limits_{\{i,j\}\in \E}\!\!\!\!  \left(\underline{c}_{ij} 
(\partial_{\theta_i}\!-\!\partial_{\theta_j})(\theta_i\!-\!\theta_j)
\!+\! D_{ij} (\partial_{\theta_i}\!-\!\partial_{\theta_j})^2(\theta_i\!-\!\theta_j)^2\right).
\end{align}}
where ${\bm \theta}=(\theta_i:i\in\V)$.
Interested to establish threshold for the M$q$P and recalling that the threshold is controlled solely by convergence of the respective integrals at the large values of $|\theta_i|$, $|\theta_i|\gg \sqrt{\kappa/D}$ (where $\kappa$ and $D$ are the largest and smallest coefficients over the network)  we ignore the $\kappa$ term in Eq.~(\ref{eq:KFP-multi}) and observe that the resulting version of the $\hat{\cal L}_{m}$ operator is scale-invariant. This suggests that the large $\theta$ asymptotic solution of the KFP equations is a scale-invariant zero modes of the $\kappa=0$ version of $\hat{\cal L}_{m}$ thus resulting in the fat, algebraic tail.  

\section{Fat Tails}\label{sec:fat-tails}

KFP analysis of the linear feedback control in the previous Section reveals that at the large values of the state vector and under assumption that a steady state is achievable the state vector's  probability distribution function shows fat algebraic tails, see e.g. Eqs.~(\ref{eq:P-sw},\ref{eq:thermal-stat}) and concluding discussion of Section \ref{sec:thermal-CSS-multi}.  Are these results, derived under the assumption of the special short-Gaussian model of the multiplicative noise, incidental or general?

In this Section,  which is arguably the main one of the letter, we utilize the general mathematical theory of stochastic processes \cite{ruelle_ergodic_1979,goldhirsch_stability_1987} to show that the fat algebraic tails are in fact general. 

\subsection{Swimmers in a general large scale flow}\label{sec:swimmers-general}

In the case of a general, smooth, large scale flow forward in time solution of Eq.~(\ref{eq:swim-r}) becomes
\begin{gather}\label{eq:swim-solution}
    {\bm r}(t)\!=\!e^{-\phi t}{\bm W}(t)\!\Big(\!\!{\bm r}(0)\!+\!\int_0^t \!dt' e^{\phi t'} {\bm W}^{-1}(t'){\bm \xi}(t')\!\!\Big),
\end{gather}
where the time-ordered exponential ${\bm W}(t)$ satisfies Eq.~(\ref{eq:W}). We will study the large time statistics of $r(t)=|{\bm r}(t)|$, where the initial separation is forgotten and thus the first term within the brackets on the right hand side of Eq.~(\ref{eq:swim-solution}) can be dropped. Following the logic of \cite{balkovsky_universal_1999} (see also references therein) we observe  that in the long-time regime, where the inter-swimmer separation $r(t)$ is significantly larger than the so-called diffusive scale $r_d= \sqrt{|\lambda|/\kappa}$ and $\lambda(t)= \max_i (\lambda_i:i=1,\cdots,d)$ is the largest (finite-time) Lyapunov exponent of the Batchelor flow, fluctuations of ${\bm r}(t)$ are mainly due to the Lyapunov exponents distributed according to Eq.~(\ref{eq:Cramer}). In other words, in this asymptotic we can approximate the inter-swimmer distance by
\begin{gather}\label{eq:swim-asymptotic}
   r(t)\approx exp\left((\lambda_1(t)-\phi)t\right) r_d. 
\end{gather}
Consistently with Eq.~(\ref{eq:Cramer}) statistics of $\lambda_1$ is governed by
\begin{gather}\label{eq:Cramer-largest}
   P(\lambda_1|t)\propto \exp\left(-t S_1(\lambda_1)\right), 
\end{gather}
where thus $S_1(\lambda_1)$ is the Cr\'{a}mer function of the largest Lyapunov exponent of the Batchelor flow achieving its minimum at $\bar{\lambda}_1$. Substituting $\lambda_1(t)$, expressed via $r(t)$ according to Eq.~(\ref{eq:swim-asymptotic}), into Eq.~(\ref{eq:Cramer-largest}) and expanding $S_1(\cdot)$ in the Taylor Series around $\bar{\lambda}_1$ we arrive at the following asymptotic expression for statistics of $r(t)$  at large but finite $t$:
\begin{align}\nonumber 
    & P(r|t) \propto {\color{black} \frac{1}{r}}\exp\left(-t S_1\left(t^{-1}\log\left(\frac{r}{r_d}\right)+\phi\right)\right)\\ 
    \nonumber & \to\! {\color{black} \frac{1}{r}}\exp\left(\!-t\!\left(\!S_1(\bar{\lambda}_1)\!+\!\left(\frac{1}{t}\log\left(\frac{r}{r_d}\right)\!+\!\phi\!-\!\bar{\lambda}_1\!\right)^2 S''_1(\bar{\lambda}_1)\right)\!\right)\\
    \label{eq:swim-general-statistics}
    & \to\Big|_{t\to\infty;\ r\gg r_d} P_{st}(r) {\color{black} \propto \frac{1}{r}}\left(\frac{r_d}{r}\right)^{2(\phi-\bar{\lambda}_1) S''_1(\bar{\lambda}_1)}.
\end{align}
Notice that the stationary version of the last line in Eq.~(\ref{eq:swim-general-statistics}) settles if ${\color{black} \phi>\bar{\lambda}_1}$ and that it is fully consistent with Eq.~(\ref{eq:P-sw}) derived for the short ($\delta$)-correlated velocity gradient,  where $S''_1(\bar{\lambda}_1)=1/((d-1)D)$ and $\bar{\lambda}_1={\color{black} (d-1) d} D/2$. 

\subsection{General Model: Synthesis}\label{sec:synthesis}
Assume that ${\bm u}(t)$ in Eq.~(\ref{eq:SDE}) is substituted by a general linear feedback $u_i(t)\to \sum_j\phi_{ij}x_j$. Then Eq.~(\ref{eq:swim-solution}) generalizes to
\begin{gather}\label{eq:x-W}
    {\bm x}(t)=e^{-({\bm m}+{\bm \phi})t}{\bm W}(t)\tilde{x},
\end{gather}
where $\tilde{x}$ stabilizes to a constant as $t$ grows. Projecting Eq.~(\ref{eq:x-W}) to the $i$-th eigen-vector 
${\bm f}_i$ of ${\bm W}$, assuming that the linear feedback is sufficiently strong and taking the $t\to\infty$ limit, we arrive at the following explicit and simple expression for the fat tail of $({\bm x}{\bm f}_i^T)$
\begin{align}
    \label{eq:multi-general-tail}
  \log P_{st}({\bm x}{\bm f}_i^T) \propto 2 {\bm f}_i\left({\bm m}+{\bm \phi}\right){\bm f}_i^T S''_i(0)\log\frac{x_d}{{\bm x}{\bm f}_i^T},
\end{align}
which is bi-linear in the sum of the system's dynamic matrix ${\bm m}$, in the linear feedback rate ${\bm \phi}$ and in the curvature $S''_i(0)$ evaluated at the minimum of the Cr\'{a}mer function of the $i$-th Lyapunov exponent of ${\bm W}$. 

Two additional remarks are in order. First, notice that dependence on $\tilde{x}$ is ``under logarithm" -- thus weak and replaced by $x_d$,  which is  an estimate for the size of the center of the probability distribution of $({\bm x}{\bm f}_i^T)$ dependent on the additive noise. Second, it follows from Eq.~(\ref{eq:multi-general-tail}) that statistics of any norm of ${\bm x}$ is equivalent to statistics of $({\bm x}{\bm f}_1^T)$ associated with the largest Lyapunov exponent $\lambda_1$.

\section{Stochastic Optimal Control (SOC)}\label{sec:SOC}

Our next step is to derive and solve Hamilton-Jacobi-Bellman (HJB) equation associated with the dynamics, governed by Eqs.~(\ref{eq:SDE}), and by the dynamic version of the cost function described by Eq.~(\ref{eq:steady-control}).  This material is auxiliary (to the main message of the letter) and it is presented here as a check of consistency and also a link to the classic (and thus benchmark) methodology in the field. Derivation of the HJB equations which follow is standard (see e.g. \cite{barber_optimal_2011} and references therein for details) thus abbreviated. 
  
We study a finite horizon version of Eqs.~(\ref{eq:SDE}), evaluated in the ``running time" $\tau$ where $\tau\in [t,t_f]$, and we introduce the cost-to-go $S(t,\x)$ considered as a function of $t$ and ${\bm x}(t)$:
\begin{gather} \label{eq:dynamic-control}
   \hspace{-0.3cm} S(t,\x)\!=\!\!\min_{\{{\bm u}(t)\}}\! \!\!\Big(\!\! S_f(\x(t_f))\!+\!\!\!\int_{t}^{t_f}\!\! \! d\tau\  \mathbb{E}\left[C(\x(\tau),{\bm u}(\tau))\right]\!\!\Big)\!,
\end{gather}
where $S_f(\x(t_f))$ is the contribution to the cost-to-go associated with the final position $\x(t_f)$; and the expectation in Eq.~(\ref{eq:dynamic-control}) is over the random ${\bm \xi}(t)$ and ${\bm \sigma}(t)$. Then $S(t,\x)$ satisfies the following HJB equation supplemented by the condition $S(t_f,\x)=S_f(\x)$:
\begin{align}\nonumber
&-\partial_t S(t,\x)  =  \min_{{\bm u}} \Big( C(\x,{\bm u})+
\left(\underline{a}_i+u_i+\underline{m}_{ij} x_j\right)\partial_{x_i}S(t,\x)\\ \label{eq:HJB}
&+\left(\kappa_{ij}\partial_{x_i}\partial_{x_j}+D_{ik;jl}x_k {\color{black}x_l}\partial_{x_i} \partial_{x_j} \right)S(t,\x)\Big).
\end{align}

\subsection{Swimmers SOC: short correlated large-scale flow}\label{sec:swimmers-SOC}

Assume that ${\bm u}$ is elongated with ${\bm r}$, i.e. ${\bm u}=u {\bm r}/r$, then $C(u,r)=u^2+\beta r^q$.  Accounting for Eqs.~(\ref{eq:simple-swimmers},\ref{eq:KFP-swimmers})  we arrive at the following "simple swimmers" version of Eq.~(\ref{eq:HJB}) 
\begin{align} \nonumber 
-\partial_t S & =\beta r^q+\frac{r^{1-d}}{2}\partial_r r^{d-1}\left(D (d-1)r^2+\kappa\right)\partial_rS+J,\\ \label{eq:HJB-swimmers} J& = 
\min_{u}\left(u^2
+ u \partial_r S\right)=-\frac{1}{4}\left(\partial_r S\right)^2,
\end{align}
where the optimal control is $u^*(t,r)=-\partial_r S(t,r)/2$. Assuming that $q=2$, function defining the final condition $S_f(r)$ is quadratic in $r$ and thus 
looking for solution of Eq.~(\ref{eq:HJB-swimmers}) in the (quadratic in $r$) form $S(t,r)=\varsigma(t)r^2+s(t)$, we arrive at the following system of equations:
\begin{align*}
    d \kappa \varsigma+ds/dt=0,\ 
    (d^2+d-2)D \varsigma-\varsigma^2+\beta+d\varsigma/dt=0,
\end{align*}
which results in the following solution for $\varsigma(t)$:
\begin{align}
    \nonumber & \varsigma(t)=\frac{1}{2}\Bigg(D(d+2)(d-1)+\sqrt{4\beta+D^2(d+2)^2(d-1)^2}\\ \label{eq:varsigma-sol} & *\tanh\left(\frac{(t_1-t)}{2}\sqrt{4\beta+D^2(d+2)^2(d-1)^2}\right)\Bigg),
\end{align}
where $t_1$ is tuned to satisfy the final condition, $S(t_f,r)=S_f(r)$. We observe that $\varsigma(t)$, defined by Eq.~(\ref{eq:varsigma-sol}) and considered at the $t\to-\infty$ asymptotic, is fully consistent with Eq.~(\ref{eq:steady-swimmers-opt}) derived under assumption of the CSS control. 

\subsection{Thermal SOC} 
\label{sec:thermal-SOC}

Formally, analysis of the single-zone version of the thermal SOC is similar to what was just described for the case of two swimmers. Adapting the HJB Eq.~(\ref{eq:HJB}) to the case of the multi-zone temperature control  and then generalizing the thermal CSS setting discussed in Section \ref{sec:thermal-CSS-multi} we arrive at the following HJB equation for the cost-to-go $S(t,{\bm\theta})$:
{\color{black}
\begin{align}\label{eq:multi-zone-HJB}  -\partial_t S  & =\!\!\!\sum\limits_{i\in \V}\!\!\!\left(\beta_i|\theta_i|^q\!+\! D_{io} \theta_i^2\partial_{\theta_i}^2
S+\kappa_i\partial_{\theta_i}^2 S\!+\!J_i\right) \\ \nonumber &
+\sum\limits_{\{i,j\}\in \E}\!\!\!\! \Big(
(\theta_i-\theta_j)(\partial_{\theta_i}-\partial_{\theta_j})\underline{c}_{ij} \\
\nonumber & + D_{ij}
(\theta_i-\theta_j)^2(\partial_{\theta_i}-\partial_{\theta_j})^2 S\Big),\\ \nonumber  
\forall i:\ & J_i =\! \min\limits_{\tilde{u}}\left(\alpha_i\tilde{u}^2\!-\!\tilde{u} c_{is}(\underline{T}-T_s)\partial_{\theta_i}S\right)\\ \nonumber  & =-\frac{\gamma_i}{2} \left(\partial_{\theta_i} S\right)^2,\ \gamma_i= \frac{c_{is}^2 (\underline{T}-T_s)^2}{2\alpha_i}.
\end{align}}
As in the other HJB examples discussed so far  we are looking for the cost-to-go, solving the multi-zone HJB Eq.~(\ref{eq:multi-zone-HJB}) correspondent to $q=2$, as a symmetric quadratic form in ${\bm \theta}$ with coefficients dependent on $t$: $S(t,{\bm \theta})=\sum_{i,j}\theta_i\varsigma_{ij}(t)\theta_j+s(t)$, where $\varsigma_{ij}=\varsigma_{ji}$. Substituting the quadratic ansatz in the Eq.~(\ref{eq:multi-zone-HJB}) we arrive at the system of generalized Ricatti equations (which we will not present here to save space). It is straightforward to check that at $\kappa=0$ and in the $t\to -\infty$ limit the system of equations is consistent with a solution of the KFP Eq.~(\ref{eq:KFP-multi}) at the optimal $\phi^{(*)}$. 

\subsection*{Conclusions and Path Forward}

We analyzed linear dynamic system driven by additive and multiplicative noise of a general position  which is stabilized by a linear feedback control. We have introduced an orthonormal basis of the time-ordered exponential of the multiplicative noise matrix and showed that stationary statistics of the state vector projected to an element of the basis shows an algebraic tail. Exponent of the tail is presented as an explicit expression (\ref{eq:multi-general-tail}) which is bi-linear in the sum of the constant part of the system's dynamic vector with the linear feedback rate matrix and in the curvature at the minimum of the Cr\'{a}mer function of the element of the basis. We believe that it is of interest to extend the approach in the future to a data driven setting where the Cr\'{a}mer functions of the Lyapunov exponents and respective eigen-vectors are learned from observations. 

Emergence of the fat tails in the linear systems driven by a multiplicative noise suggests using the newly introduced Mean-$q$-Power criterion for adjusting the linear feedback control. We have illustrated that the criterion can be validated on and is useful for approaches which are classic in control, specifically for Control of Steady State (Section \ref{sec:CSS}) and for Stochastic Optimal Control (Section \ref{sec:SOC}). We plan to extend this approach to systems with partial observability and control-dependent noise. 

Our results are motivated by and illustrated on examples from hydrodynamics and civil engineering. We envision extending this work to more complex fluid flows and "white box" modeling of the thermostatically control multi-zone buildings. This can be achieved via  data driven reinforcement learning approaches improving the control schemes described in this letter \footnote{The author is grateful to L. Pagnier, R. Ferrando, C. Koh, S Konkimalla and A. Larsen for multiple discussions.  This work is a part of the team collaboration.}.

\bibliographystyle{IEEEtran}

\bibliography{bib/extra,bib/ActiveMatterFieldsParticles,bib/SmoothVelocity,bib/StochasticOptimalControl,bib/TCL,bib/NetworkControl-MultiplicativeDelays}

\end{document}